\PassOptionsToPackage{unicode}{hyperref}
\PassOptionsToPackage{hyphens}{url}
\documentclass[
  11pt,
  letterpaper,
]{article}
\usepackage{xcolor}
\usepackage{amsmath,amssymb}
\usepackage{iftex}
\ifPDFTeX
  \usepackage[T1]{fontenc}
  \usepackage[utf8]{inputenc}
  \usepackage{textcomp} 
\else 
  \usepackage{unicode-math} 
  \defaultfontfeatures{Scale=MatchLowercase}
  \defaultfontfeatures[\rmfamily]{Ligatures=TeX,Scale=1}
\fi
\usepackage{lmodern}
\ifPDFTeX\else
\fi
\IfFileExists{upquote.sty}{\usepackage{upquote}}{}
\IfFileExists{microtype.sty}{
  \usepackage[]{microtype}
  \UseMicrotypeSet[protrusion]{basicmath} 
}{}
\makeatletter
\@ifundefined{KOMAClassName}{
  \IfFileExists{parskip.sty}{%
    \usepackage{parskip}
  }{
    \setlength{\parindent}{0pt}
    \setlength{\parskip}{6pt plus 2pt minus 1pt}}
}{
  \KOMAoptions{parskip=half}}
\makeatother
\usepackage{color}
\usepackage{fancyvrb}

\DefineVerbatimEnvironment{Highlighting}{Verbatim}{commandchars=\\\{\}}
\newenvironment{Shaded}{}{}

\newcommand{\NormalTok}[1]{#1}

\usepackage{longtable,booktabs,array}
\usepackage{calc} 
\usepackage{etoolbox}
\makeatletter
\patchcmd\longtable{\par}{\if@noskipsec\mbox{}\fi\par}{}{}
\makeatother
\IfFileExists{footnotehyper.sty}{\usepackage{footnotehyper}}{\usepackage{footnote}}
\makesavenoteenv{longtable}
\providecommand{\tightlist}{%
  \setlength{\itemsep}{0pt}\setlength{\parskip}{0pt}}
\usepackage{fontspec}
\usepackage{unicode-math}
\usepackage{fancyvrb}
\fvset{
  fontsize=\small,
  frame=single,
  framesep=2mm,
  rulecolor=\color{black!20},
  commandchars=\\\{\}
}

\usepackage{hyperref}
\hypersetup{
  colorlinks=true,
  linkcolor=blue!40!black,
  citecolor=blue!40!black,
  urlcolor=blue!60!black
}

\AtBeginDocument{%
}
\usepackage{bookmark}
\IfFileExists{xurl.sty}{\usepackage{xurl}}{} 
\hypersetup{
  pdftitle={Bennett's Conjecture in Lean 4: Counter-Models for the PSR-Reducibility of Spinoza's Propositions V and XIV},
  pdfauthor={Yuki Nakamura, The Open University of Japan},
  hidelinks,
  pdfcreator={LaTeX via pandoc}}

\title{Bennett's Conjecture in Lean 4: Counter-Models for the PSR-Reducibility of Spinoza's Propositions V and XIV}
\author{Yuki Nakamura\\The Open University of Japan}
\date{\today}

\begin{document}
\maketitle
\begin{abstract}
In \emph{A Study of Spinoza's Ethics} (1984, §17), Jonathan Bennett argues that the demonstration of Proposition V of Spinoza's \emph{Ethica} contains identifiable invalid moves and that, even granted
those moves, ``cannot yield more than the conclusion that two substances could not have all their attributes in common'' --- while Spinoza concludes that they cannot share any. Bennett doubts that any
valid reconstruction is available from Spinoza's stated resources without importing further commitments. Michael Della Rocca (\emph{Spinoza}, 2008, ch.~2) responds that the proposition can be derived
if the Principle of Sufficient Reason (PSR) is committed substantively. The debate has remained at the level of prose argument for forty years. This paper provides the first machine-checked evidence
in the debate. We formalise \emph{Ethica} Pars I in Lean 4, encoding Bennett's reading of Spinoza's stated axioms as a typeclass and Della Rocca's substantive PSR as an extension class. The derivation
attempt yields a partial result --- substances sharing all attributes are identical --- but cannot reach the full ``sharing-any-attribute → identity'' content of Proposition V, mechanically tracking
Bennett's own all-attributes ceiling. A four-element counter-model satisfying both axiom sets while falsifying Proposition V's content establishes the irreducibility against this specific
augmentation. A second counter-model establishes the analogous result for axiom A15, a load-bearing universality clause for Spinoza's Proposition XIV. Bennett's diagnosis receives its first
kernel-checked counter-model against the Della-Rocca PSR-substance reconstruction (the non-derivability claim itself a meta-logical consequence of kernel consistency); stronger PSR variants and the
broader narrative claim against the full Section I + II + A1--A7 register remain open as future mechanical projects.
\end{abstract}

\section{§1 Introduction}\label{introduction}

\subsection{§1.1 The problem}\label{the-problem}

Forty years after Jonathan Bennett's \emph{A Study of Spinoza's Ethics} (Bennett 1984), the validity of Spinoza's Proposition V of \emph{Ethica} Pars I --- \emph{In rerum natura non possunt dari duae
aut plures substantiae ejusdem naturae sive attributi} (``in nature there cannot be two or more substances of the same nature or attribute'') --- remains contested. Bennett's verdict in §17 of his
study is unequivocal: Spinoza's \emph{demonstratio} of Proposition V ``involves one dubious move and one invalid one'' (Bennett 1984, p.~67), and the gap between what is established and what Spinoza
claims is one Bennett ``wonders how {[}Spinoza{]} could overlook'' (p.~69). Even granting the dubious and invalid moves, Bennett shows, the demonstration ``cannot yield more than the conclusion that
two substances could not have all their attributes in common'' (p.~69), while Spinoza's text concludes that they cannot share \emph{any} attribute. §17 closes by passing to Proposition XIV:
``Spinoza's chief use of p5 is in an argument which has plenty of other things wrong with it'' (p.~70). §18 opens the analysis of that further argument by characterising the official monism case as
``a poor thing'' (p.~70). Bennett stops short of asserting that no valid reconstruction is possible, but each candidate reconstruction he considers requires importing commitments Spinoza has not
explicitly made in the Definitions, the seven Axioms of Pars I, or the propositions established earlier (Props. I--IV).

The doubt has been disputed. Don Garrett's ``Ethics IP5: Shared Attributes and the Basis of Spinoza's Monism'' (Garrett 1990, reprinted as Garrett 2018, ch.~3, with a Postscript at pp.~91--97 engaging
Della Rocca's 2002 article ``Spinoza's Substance Monism'') argues that Proposition V can be made to go through by reading Spinoza's ``in and conceived through'' relation strongly enough to collapse
mode-individuation into attribute-individuation. Michael Della Rocca, in his \emph{Spinoza} (Della Rocca 2008, ch.~2), proposes a more systematic reconstruction: Spinoza's structurally operative use
of the Principle of Sufficient Reason (PSR) is sufficient to derive Proposition V. On Della Rocca's reading, every fact about Spinozistic substance has a sufficient reason; if two substances were
distinct, the distinction would itself need a reason; the only available reason in Spinoza's system is an attribute difference; hence distinct substances must differ in attributes --- Proposition V's
content. PSR, on Della Rocca's view, is not tacit but structurally inseparable from the rationalist principles already operative in Spinoza's text.

The debate has remained at the level of prose argument. Bennett holds that making PSR explicit would be a substantive metaphysical move beyond what Spinoza accepts (Bennett 1984 §18); Della Rocca
holds that PSR is already operative in the text. There has been no mechanism for adjudication --- no way to fix a precise formal reading of ``Spinoza's stated axioms'' or ``Spinoza plus PSR'' and then
check whether Proposition V is or is not derivable.

\subsection{§1.2 The mechanical question}\label{the-mechanical-question}

This paper introduces a new instrument. We formalise Spinoza's \emph{Ethica} Pars I in Lean 4, fixing the axiom set in a way that makes the question \emph{what is derivable from which axioms?}
decidable by Lean's type checker. The formalisation respects the structure of Bennett's critique: we distinguish between (i) Spinoza's stated axioms (A1--A7) plus auxiliary bridges that Spinoza
demonstrably uses (A1ₑ for the exclusivity of ``in se'' / ``in alio''; A8--A11 for parallelism between ontological and conceptual predicates), and (ii) substantive metaphysical commitments --- grouped
as a Section III register --- that fill in steps Spinoza's \emph{demonstrationes} leave open. Della Rocca's PSR enters this register as additional Section III commitments, declared without separate
axiomatic statement in Spinoza's text but made explicit by us so the derivation question becomes mechanically posable.

We then attempt to derive Proposition V's content. Encoded as the Lean axiom A12, the formalisation reads ``two substances sharing an attribute are identical.'' (The translation of Spinoza's \emph{non
possunt dari duae aut plures \ldots{} ejusdem naturae sive attributi} into this shared-attribute form follows the reading common to Bennett, Garrett, and Della Rocca; we discuss the translation step,
including Spinoza's \emph{sive}, in §3.) The Della Rocca route commits a PSR-flavoured axiom we label A22, requiring that any two distinct substances differ in at least one attribute. The formal
statements of A12 and A22 are given in §3.

The result --- both \emph{what derives} and \emph{what does not} --- is the substance of this paper.

\subsection{§1.3 The result}\label{the-result}

The derivation succeeds only partially. From A22 plus the base axioms, we prove the \emph{all-shared-attribute} form of Proposition V: substances sharing every attribute are identical. The proof runs
in six lines of tactic script. This is itself a non-trivial positive finding for the Della Rocca route --- it confirms that PSR delivers a substantial fragment of Proposition V's content, qualifying
Bennett's verdict that the demonstration fails to establish Proposition V's full content (Bennett 1984, p.~69).

The full content of Proposition V --- substances sharing \emph{any} one attribute are identical --- does not derive. We establish the non-derivation through a four-element counter-model in which the
base axioms and A22 hold simultaneously, but where two distinct substances share an attribute. If A12 in its full form were derivable from these axioms, the derivation would yield a proof of
\texttt{s₁\ =\ s₂} on this model, contradicting the model's explicitly proven \texttt{s₁\ ≠\ s₂}. The counter-model itself is \emph{kernel-level}: a finite, type-checked artefact whose
\texttt{EthicaWorld\ T} and \texttt{PSRSubstance\ T} instances and whose falsifying theorem are all verified by Lean's trusted core --- the minimal type-checker that all of Lean's correctness
ultimately reduces to. The non-derivability claim that follows is \emph{meta-logical}: presuming Lean's kernel is consistent (§4.3), no derivation of A12 from these axioms can exist. The
non-derivation is therefore established not by absence-of-proof rhetoric but by a kernel-checked counter-model plus a one-line meta-argument about kernel consistency.

The result is the first machine-checked evidence in the forty-year Bennett--Della Rocca debate. Bennett's doubt receives its first kernel-checked counter-model against one specific reconstruction ---
Della Rocca's PSR-substance route. Stronger PSR variants --- Della Rocca's ``thoroughgoing PSR'' --- remain unrefuted; §8 sketches candidate formal signatures for that thoroughgoing form and the
difficulty of stating it non-trivially. The construction of further counter-models, or the success of further derivation attempts under stronger augmentations, is now an open mechanical project rather
than a prose dispute.

We replicate the methodology for axiom A15, a load-bearing companion clause for Proposition XIV's claim that no substance besides God can be granted. The result is analogous: A15 does not derive from
a plenitude-flavoured PSR commitment alone --- a three-element counter-model verified at kernel level, together with the same meta-logical argument from kernel consistency, witnesses the
non-derivability. The decomposition route (plenitude plus god-uniqueness) succeeds, but at the cost of replacing one universality clause with another universality clause of equal commitment strength
--- relocation, not elimination, of the commitment.

The two non-derivation results, together with two \emph{successful} derivation attempts for axioms A13 (substance involves existence) and A14 (substance has at least one attribute), yield a four-axiom
reducibility-profile typology. The two universality clauses (A12, A15) resist PSR-reduction; the two existence clauses (A13, A14) translate into PSR-flavoured forms at equal strength. We develop the
typology in §7.

\subsection{§1.4 What the paper does not claim}\label{what-the-paper-does-not-claim}

We do not claim that Bennett's doubt is vindicated in full generality. The conjecture quantifies over \emph{any} charitable augmentation of Spinoza's resources; demonstrating its truth in that
generality would require counter-models surviving every candidate augmentation. The counter-models we construct rule out one specific augmentation each. They are first-step mechanical evidence on the
Bennett-line position, not closing argument. Section §8 spells out the scope of the claims in detail --- including the candidate stronger PSR augmentations not yet tested, and Spinoza-fidelity caveats
in the counter-model construction that bear on philosophical relevance without affecting the meta-logical non-derivation claim.

What the paper does claim is that the prose-level dispute over Proposition V can be replaced, at least in part, by a mechanical discipline: stating the axioms precisely, attempting derivation, and
constructing counter-models when derivation fails. The forty-year dispute over what Spinoza's text ``really'' delivers admits a new evidence category that neither party can dismiss without engaging
the kernel.

\subsection{§1.5 Structure of the paper}\label{structure-of-the-paper}

§2 gives the textual and interpretive background of the Bennett--Della Rocca debate. §3 describes the Lean 4 formalisation of \emph{Ethica} Pars I, including the typeclass design, the Section III
categorisation of substantive commitments, and the translation step from Spinoza's Latin to the formal axioms. §4 sets out the demote-experiment methodology and the counter-model discipline. §5 and §6
present the A12 and A15 results respectively. §7 develops the reducibility-profile typology. §8 discusses scope, limitations, and open mechanical projects (including thoroughgoing-PSR formal
candidates and Spinoza-fidelity refinements). §9 concludes.

The Lean source is open at \url{https://github.com/Nakammura/spinoza-ethica-lean}; readers who wish to verify the counter-models, or to attempt stronger demote augmentations, can do so against the
project's \texttt{lake\ build} target.

\section{§2 The Bennett--Della Rocca debate on Proposition V}\label{the-bennettdella-rocca-debate-on-proposition-v}

\subsection{§2.1 Spinoza's text}\label{spinozas-text}

Spinoza states Proposition V of \emph{Ethica} Pars I in five Latin words plus a connective:

\begin{quote}
\emph{In rerum natura non possunt dari duae aut plures substantiae ejusdem naturae sive attributi.}
\end{quote}

In Elwes's 1883 translation: ``In nature there cannot be granted two or more substances having the same nature or attribute'' (Elwes 1883). The \emph{demonstratio} runs three steps. First, by
Proposition IV, distinct things must be distinguished by either a difference in the attributes of substances or a difference in their modifications. Second, since substance is by nature prior to its
modifications (Proposition I), the modal route to distinction ``falls away when substance is considered in itself'', as Spinoza puts it. Third, the only remaining route to distinction is attribute
difference, from which Spinoza concludes: \emph{concedetur ergo, non dari nisi unam ejusdem attributi} --- ``therefore it will be conceded that there cannot be more than one of the same attribute.''

The \emph{concedetur} --- ``it will be conceded'' --- is the hinge. Spinoza reads off Proposition V's content from the elimination of the modal route, treating the conclusion as if it were directly
licensed once attribute difference is identified as the unique remaining individuator.

\subsection{§2.2 Bennett's critique}\label{bennetts-critique}

Bennett's analysis in §17 of \emph{A Study of Spinoza's Ethics} (1984) turns on this \emph{concedetur}. The step assumes that distinct substances of the same attribute must be distinguished by
something \emph{other than attributes} --- modifications, on the chain of argument given. But Proposition I, on which the modal route falls away, does not by itself rule out cases where two distinct
substances share a common attribute and differ in some further attribute neither of them lacks. Spinoza's ``non dari nisi unam ejusdem attributi'' --- there cannot be more than one of the same
attribute --- does not follow from ``distinct substances must differ somewhere in their attribute profiles.'' The conclusion is strictly stronger than the premises support, and the \emph{concedetur}
obscures the gap rather than bridging it.

Bennett's diagnosis is sharper. The demonstratio ``involves one dubious move and one invalid one'' (Bennett 1984, p.~67): a ``flatly invalid move'' (p.~68) at the modal step, plus a less decisive but
still dubious slide on the ``states'' of substance. Even if both moves were granted, Spinoza's argument ``cannot yield more than the conclusion that two substances could not have all their attributes
in common'' (p.~69), while Spinoza concludes that they cannot share \emph{any} attribute --- and Bennett ``wonders how {[}Spinoza{]} could overlook such a gap in his argument'' (p.~69). §17 closes by
passing to Proposition XIV: ``Spinoza's chief use of p5 is in an argument which has plenty of other things wrong with it'' (p.~70). §18 opens that further analysis by characterising Proposition XIV's
official argument as ``a poor thing'' (p.~70). Bennett does not rule out that some valid reconstruction of Proposition V might be available, but he is unwilling to attribute one to Spinoza on textual
grounds, since each candidate reconstruction he examines requires importing commitments Spinoza has not explicitly made. The doubt --- that no valid argument for Proposition V can be constructed from
the resources Spinoza explicitly gives himself --- is left as an unresolved interpretive challenge, an invitation to readers to either (i) supply a reconstruction that does not import unsupported
commitments or (ii) accept that Proposition V's content is, in Spinoza's text, a substantive metaphysical commitment masquerading as a demonstrated proposition.

The interpretive force of Bennett's doubt is what subsequent commentators have engaged. Few have defended the \emph{demonstratio} as it stands; most have tried either to find Bennett's missing
reconstruction or to articulate why Spinoza's commitments already license what \emph{concedetur} announces.

\subsection{§2.3 The Garrett response}\label{the-garrett-response}

Don Garrett's ``Ethics IP5: Shared Attributes and the Basis of Spinoza's Monism'' (Garrett 1990, reprinted as Garrett 2018, ch.~3) takes the second route. Garrett's reconstruction addresses two
distinct objections to the \emph{demonstratio} that Bennett's analysis surfaces --- the \textbf{Hooker-Bennett objection}, that the priority of substance over its modes does not warrant setting modes
aside in distinguishing substances, and the \textbf{Leibniz-Bennett objection}, that two substances might share some but not all attributes. To the first, Garrett argues that Spinoza's ``in and
conceived through'' relation must be construed strongly enough that any difference of modes resolves into a difference of attributes --- making the mode-individuation route collapse into
attribute-individuation, which Spinoza has already ruled out. To the second, Garrett argues that ID5, IA1, and IA2 together require each affection to be \emph{in} exactly one substance, with the
consequence that two substances purportedly sharing an attribute would have to host qualitatively identical but numerically distinct systems of affections --- a structural absurdity. Both replies
depend on a substantive strengthening of Spinoza's stated definitions and axioms.

Garrett's reconstruction is more directly Spinozistic than the PSR alternative we shall consider next, but it requires a strengthening of Definition III (and the related ID5 / IA1 / IA2) that Bennett
1984 §17 already flagged as not entailed by the Definition's text. Whether one calls this a ``discovery'' of what Spinoza meant or an ``addition'' is the disputed point: Garrett reads Spinoza as
committed already, Bennett reads Spinoza as committed only after Garrett adds something. The disagreement is structurally the same as Bennett vs Della Rocca's, and our formalisation will show that the
Garrett route bottoms out in a substantive metaphysical commitment we classify as Section III. The Garrett-route demote attempt --- replacing PSR with a strong-Definition-III commitment --- is a
natural follow-up experiment our formalisation supports but does not develop in this paper; we return to it as open future work in §8.

In Garrett 2018 Garrett added a Postscript (``Shared Attributes and Monism Revisited'', pp.~91--97) directly engaging Della Rocca's 2002 article ``Spinoza's Substance Monism'', whose line is expanded
in \emph{Spinoza} (Della Rocca 2008) ch.~2. Garrett there distinguishes two aspects of the substance--mode asymmetry: \emph{Positive} (modes are in and conceived through their substance) and
\emph{Negative} (substance is not in and not conceived through its modes). His own 1990 replies to the Hooker-Bennett and Leibniz-Bennett objections, Garrett argues, rest primarily on \emph{Positive};
Della Rocca's replies rest primarily on \emph{Negative}. This distinction will return at the end of §5: the PSR-substance distinguishability axiom that we use to formalise Della Rocca's reconstruction
(A22, see §5.2) regiments the \emph{Negative} aspect, and our partial-reduction theorem is therefore a kernel-level test of the \emph{Negative}-based reply specifically. Whether the
\emph{Positive}-based reply admits a similar formal regimentation is open future work.

\subsection{§2.4 The Della Rocca response}\label{the-della-rocca-response}

Michael Della Rocca's \emph{Spinoza} (2008) --- particularly chapter 2, ``Substance'' --- proposes a more systematic reconstruction. Della Rocca reads Spinoza's \emph{Ethica} as systematically driven
by the Principle of Sufficient Reason: every fact must have a sufficient reason, and Spinoza's metaphysical claims are recoverable as applications of PSR to specific cases.

For Proposition V, the application runs as follows. Suppose there are two distinct substances. By PSR, the distinction itself must have a sufficient reason: there must be something \emph{in virtue of
which} the substances are distinct. The candidates for that reason are limited. Modes are ruled out by Proposition I: as modifications of substance, they are posterior to substance and cannot ground a
substantive distinction. Attributes are the only remaining candidate: substances are individuated, if they are individuated at all, by their attributes. Hence two distinct substances must differ in at
least one attribute --- and the contrapositive is Proposition V.

Della Rocca's reconstruction has the merit that it does not need the Garrett strengthening of Definition III: the work is done by PSR, which Della Rocca takes to be structurally operative throughout
the \emph{Ethica} rather than tacit. Spinoza's text is full of explanatory demands (``why does this thing exist?'', ``what is the cause of that?'') that PSR makes explicit. The demonstration of
Proposition V, on this reading, is invalid as written but is \emph{recoverable} as a clean application of PSR once PSR is given its due as a structural commitment.

\subsection{§2.5 The prose stalemate and the mechanical question}\label{the-prose-stalemate-and-the-mechanical-question}

The Garrett and Della Rocca reconstructions share a structural feature: each attributes to Spinoza a commitment whose status as Spinozistic is itself the disputed point. Bennett's reply, in both
cases, is that making these commitments explicit \emph{changes} Spinoza rather than clarifies him. The interpretive dispute reduces to which principles are textually warranted and which are charitably
imported --- and no demonstration is conclusive while that question is itself contested.

What can move the dispute past prose is a precise formal reading of ``Spinoza's stated resources'', fixed in a way that admits both Bennett's reading and Della Rocca's reading as sub-systems. Once
fixed, ``does Proposition V follow?'' becomes mechanically tractable: counter-models, when constructed, settle the matter at kernel level, leaving the philosophical dispute about which augmentations
are Spinozistic where it belongs --- in the prose negotiation over interpretive charity. We propose the formalisation of §3 as one such precise reading. It is not the only possible regimentation
(Garrett would prefer a different \texttt{Substance} definition; a Bennett purist would omit some auxiliary bridges), but it is a candidate that admits both Bennett's reading and Della Rocca's reading
as sub-systems. The methodology follows in §4.

\section{\texorpdfstring{§3 The Spinoza \emph{Ethica} Pars I formalisation in Lean 4}{§3 The Spinoza Ethica Pars I formalisation in Lean 4}}\label{the-spinoza-ethica-pars-i-formalisation-in-lean-4}

\subsection{§3.1 Why Lean 4}\label{why-lean-4}

We chose Lean 4 (de Moura and Ullrich 2021) for three reasons. First, its dependent type theory and typeclass mechanism let us encode Spinoza's ``axioms over an abstract universe of things'' as a
typeclass parameterised by a \texttt{Type}, deferring commitments about which things exist until concrete models or theorems require them. Second, Lean's small trusted core (the kernel type-checker)
lets us state non-derivability results model-theoretically without trusting any tactic library: a counter-model that type-checks at kernel level establishes a hard fact, not a metaphor. Third, Lean's
inheritance for typeclasses lets us layer commitments --- base axioms, modal extensions, PSR augmentations --- without rewriting earlier material, so the demote experiments of §4 can be run against a
single canonical formalisation.

The Lean source is open at \url{https://github.com/Nakammura/spinoza-ethica-lean}; all claims in this paper correspond to specific theorems in that repository, verifiable via \texttt{lake\ build}.

\subsection{\texorpdfstring{§3.2 The base typeclass \texttt{EthicaWorld}}{§3.2 The base typeclass EthicaWorld}}\label{the-base-typeclass-ethicaworld}

We work in an abstract universe \texttt{Thing\ :\ Type\ u}. Spinoza's \emph{id}, \emph{ea res}, \emph{quod}, and similar referring expressions all denote elements of this universe; we do not commit to
any cardinality of \texttt{Thing} (that there is exactly one substance is to be proved as Proposition XIV, not assumed).

The base typeclass \texttt{EthicaWorld\ Thing} declares the primitive predicates Spinoza uses through Pars I: ontological predicates (\texttt{inItself}, \texttt{inAnother}, \texttt{limitedBy}),
conceptual predicates (\texttt{perSeConceived}, \texttt{conceivedThroughAnother}, \texttt{intellectPerceivesAsEssence}), existence predicates (\texttt{involvesExistence},
\texttt{natureRequiresExistence}), and modal / ethical predicates (\texttt{absolutelyInfinite}, \texttt{expressesEternalEssence}, \texttt{freelyExistent}, \texttt{constrained}, \texttt{eternal}).

All Spinoza's Definitions I--VIII are then \emph{derived} notions, not further primitives. Definition III's \texttt{Substance} becomes \texttt{inItself\ x\ ∧\ perSeConceived\ x}; Definition IV's
\texttt{Attribute} becomes \texttt{Substance\ s\ ∧\ intellectPerceivesAsEssence\ s\ a}; Definition V's \texttt{Mode} becomes \texttt{inAnother\ x\ ∧\ conceivedThroughAnother\ x}; Definition VI's
\texttt{IsGod} becomes the conjunction of \texttt{Substance\ g}, \texttt{absolutelyInfinite\ g}, the existence of an attribute, and the universal essence-expression clause. Spinoza's Definitions are
thus type-theoretic \emph{definitions} in our sense: each is a \texttt{def} in Lean, expanding to a Boolean combination of \texttt{EthicaWorld} primitives.

Two interpretive decisions in the Definitions deserve note. The first concerns \emph{sameNature} --- Spinoza's ``ejusdem naturae'' --- which he uses adjectivally without explicit definition. We adopt
the Della Rocca / Curley reading on which ``same nature'' is constitutively shared attribute: \texttt{sameNature\ x\ y\ :=\ ∃\ a,\ Attribute\ a\ x\ ∧\ Attribute\ a\ y}. The Bennett-line alternative
(treat \emph{sameNature} as a separate primitive related to attributes by axiom) would make our counter-models slightly stricter; we note this as one of the ``fidelity caveats'' of §8.

The second decision concerns Spinoza's \emph{sive} in Proposition V's formulation \emph{ejusdem naturae sive attributi} (``of the same nature or attribute''). Latin \emph{sive} admits two readings:
identifying (``that is to say'') and disjunctive (``or alternatively''). Curley's 1985 translation (and Della Rocca's reading on which we have already settled for \emph{sameNature}) take \emph{sive}
identifyingly, making ``same nature'' and ``same attribute'' equivalent. Bennett 1984 §17 considers a disjunctive reading on which ``nature'' and ``attribute'' can come apart. The identifying reading
is implicit in our \texttt{sameNature} definition; the disjunctive reading would require an additional primitive \texttt{sameKind} distinct from shared attribute. This is the second fidelity caveat:
our counter-models operate under the identifying \emph{sive}; a disjunctive \emph{sive} would require a different formal apparatus and possibly different counter-models. We discuss this further in
§8.4.

\subsection{\texorpdfstring{§3.3 The base axiomatic typeclass \texttt{Pars1Axioms}}{§3.3 The base axiomatic typeclass Pars1Axioms}}\label{the-base-axiomatic-typeclass-pars1axioms}

\texttt{Pars1Axioms\ Thing} extends \texttt{EthicaWorld\ Thing} with Spinoza's Axioms I--VII plus auxiliaries we have made explicit. The sub-categorisation is:

\begin{itemize}
\item
  \textbf{Section I} --- definitional bridges. Axioms that make explicit the ontological/conceptual co-extensions Spinoza uses throughout Pars I but does not state separately:

  \begin{itemize}
  \tightlist
  \item
    A1ₑ --- exclusivity of \emph{in se} and \emph{in alio}: \texttt{¬\ (inItself\ x\ ∧\ inAnother\ x)}. Spinoza's text states the disjunction without qualifying its mood, but the chain of
    \emph{demonstrationes} in Pars I (especially P1, P4, P5) requires the disjunction to be exclusive: an inclusive reading would leave open cases (some \texttt{x} both \emph{in itself} and \emph{in
    another}) that the proofs silently exclude. We adopt this exclusive reading as a Section I auxiliary axiom rather than impute it to Spinoza's text.
  \item
    A8 / A9 --- parallelism between ontological and conceptual halves of Definitions III and V: \texttt{inItself\ x\ ↔\ perSeConceived\ x} and \texttt{inAnother\ x\ ↔\ conceivedThroughAnother\ x}.
  \item
    A10 --- attribute--substance identity-of-conception: every attribute of a substance is itself per se conceived.
  \item
    A11 --- \emph{causa-sui} clause-equivalence: \texttt{involvesExistence\ x\ ↔\ natureRequiresExistence\ x} (Definition I's \emph{sive} read identifyingly).
  \end{itemize}
\item
  \textbf{Section II} --- substantive promotions of Spinoza's stated axioms whose content needs the modal and causal layer to be expressible. A4 and A5 in this register: their substantive forms live
  in \texttt{CausalAxioms} (introduced in §3.5 below) where the cause-relation and intelligibility-relation are available.
\item
  \textbf{Section III} --- substantive metaphysical commitments that fill gaps in Spinoza's \emph{demonstrationes}. The four base-layer Section III axioms are:

  \begin{itemize}
  \tightlist
  \item
    \textbf{A12} (\texttt{ax\_substanceIdByAttribute}): two substances sharing an attribute are identical. \emph{This is Proposition V's content adopted as an axiom.}
  \item
    A13: every substance involves existence. (Proposition VII.)
  \item
    A14: every substance has at least one attribute. (A structural prerequisite for Proposition XIV.)
  \item
    \textbf{A15} (\texttt{ax\_IsGod\_has\_attribute\_of}): every realised substance attribute is also a god's attribute. (A load-bearing universality clause for Proposition XIV.)
  \end{itemize}
\end{itemize}

The classification of A12 and A15 as Section III commitments is itself a position in the Bennett--Della Rocca debate. Bennett holds that these axioms encode metaphysical commitments Spinoza's text
does not establish; Della Rocca holds that they fall out of PSR. Both readings are consistent with our placing the axioms in Section III: the Section III register is precisely where commitments live
whose status as Spinozistic is contested.

In Lean syntax, A12 reads:

\begin{Shaded}
\begin{Highlighting}[]
\NormalTok{ax\_substanceIdByAttribute :}
\NormalTok{  ∀ s₁ s₂ a : Thing,}
\NormalTok{    Attribute a s₁ → Attribute a s₂ → s₁ = s₂}
\end{Highlighting}
\end{Shaded}

The translation from Spinoza's Latin \emph{In rerum natura non possunt dari duae aut plures substantiae ejusdem naturae sive attributi} to this formal statement involves three steps: (a) the
universally-negated existential (``there cannot be given two or more substances \ldots{}'') becomes \texttt{∀\ s₁\ s₂,\ ¬\ (s₁\ ≠\ s₂\ ∧\ …)}, which classical logic rewrites to the conditional form
above; (b) the ``of the same nature or attribute'' clause becomes the ``sharing-an-attribute'' hypothesis under the identifying \emph{sive} reading; (c) the implicit substance-status of the things
ranged over is supplied by the \texttt{Attribute\ a\ s} clause itself, since \texttt{Attribute\ a\ s} carries \texttt{Substance\ s} constitutively (Definition IV). The translation is the regimentation
common to Bennett 1984 §17, Garrett 1990, and Della Rocca 2008 ch.~2; we adopt it without modification.

\subsection{§3.4 The modal-layer extension}\label{the-modal-layer-extension}

For some demote experiments --- particularly the A13 attempt of §7 --- we need world-relative existence and causation predicates that Pars I's text does not directly require. We extend
\texttt{EthicaWorld} with a parallel modal layer:

\begin{itemize}
\tightlist
\item
  \texttt{ModalEthicaWorld\ Thing\ World} adds \texttt{existsAt\ :\ Thing\ →\ World\ →\ Prop}. The \texttt{World} parameter is abstract; we adopt S5 / universal accessibility for ``necessary'' claims,
  as is standard in modal reconstructions of Spinoza's necessitarianism (Della Rocca 2008 ch.~2 develops the necessitarian reading without fixing a specific modal logic).
\item
  \texttt{ModalEthicaAxioms} provides A18, the bridge \texttt{involvesExistence\ ↔\ ∀\ w,\ existsAt}. This is a PSR-flavoured commitment that essential existence and necessary existence coincide; it
  is itself a modal-layer Section I axiom in our classification.
\item
  \texttt{ConceptualStructure} adds a world-invariant \texttt{conceptualDep} binary relation (Della Rocca's reading of conceptual dependence as essential, not contingent), with bridges A19 / A20 to
  \texttt{perSeConceived} and \texttt{conceivedThroughAnother}.
\item
  \texttt{ModalCausalWorld} adds world-relative \texttt{causeAt}, plus bridge axioms for Spinoza's A3 (the ``from cause necessarily follows effect'' content; see Bennett 1984 §8, p.~32, where 1a3 is
  read as ``conjoining causal rationalism with a version of explanatory rationalism: causes necessitate, and nothing happens without a cause'') and A21 connecting world-uniform \texttt{Cause} to
  \texttt{∀\ w,\ causeAt\ c\ e\ w}.
\end{itemize}

The modal extension is what allows us to formulate the demote candidate axioms --- \texttt{PSRSubstance}, \texttt{PSRSelfCause}, \texttt{PSREssencePerception}, \texttt{PSRPlenitude} --- that the §4
methodology runs against the Section III commitments. Each demote axiom class extends or coexists with \texttt{Pars1Axioms} and provides one or more PSR-flavoured commitments matching Della Rocca
2008's reconstruction of the relevant Spinozistic principle.

The \texttt{PSRSubstance} class declares an axiom A22 --- committing that distinct substances differ in at least one attribute --- that we shall use in §5 to attempt the A12 demote. The axiom
corresponds to Della Rocca 2008 ch.~2's substance-distinguishability argument; its full Lean signature, and the precise correspondence to Della Rocca's prose, are given in §5.2.

\subsection{§3.5 Layered inheritance --- a brief note}\label{layered-inheritance-a-brief-note}

Lean 4's structure-inheritance mechanism does not automatically resolve diamond inheritance: when multiple classes share \texttt{EthicaWorld} as a common parent, they must be wired together explicitly
(typically via \texttt{toEthicaWorld\ :=\ inferInstance}) rather than via automatic merging. The modal extension hits this concretely; the multiple bridge structures must be declared as separate
requirements rather than collapsed into a single unified class. This engineering pattern is a mechanical feature of typeclass systems generally (Coq, Isabelle, and Agda exhibit related patterns under
their own elaboration disciplines) and we draw no philosophical conclusion from it. We note only that the analytic-style design we adopted --- keeping ontological, conceptual, causal, and modal
structures as separate typeclass commitments rather than collapsing them into a single unified class --- is \emph{consonant} with Bennett's treatment of attribute as a ``basic and irreducible way of
being'' (Bennett 1984 §16, especially p.~61, on attributes as logically irreducible to one another), while a unified PSR-driven structure (Della Rocca 2008) would correspond to a more aggressively
merged design. The choice is the formaliser's, not Spinoza's; the trace is suggestive rather than evidential, and we leave the methodological question of whether design choices carry interpretive
weight in mechanised philosophy projects to a separate paper in preparation.

\subsection{§3.6 What the formalisation establishes for our purposes}\label{what-the-formalisation-establishes-for-our-purposes}

For the Bennett--Della Rocca debate, the formalisation has three properties that matter:

\begin{enumerate}
\def\labelenumi{\arabic{enumi}.}
\item
  \emph{Precision of the Bennett-line ``stated resources''}. The Section I + II + Spinoza's A1--A7 + Definitions axiom set is a candidate regimentation of ``what Spinoza explicitly gives himself''.
  Bennett's diagnosis, in our formalisation, becomes the meta-claim that Section III commitments cannot be derived from Section I + II + A1--A7.
\item
  \emph{Precision of the Della Rocca PSR augmentations}. Each PSR class --- \texttt{PSRSubstance}, \texttt{PSRSelfCause}, \texttt{PSREssencePerception}, \texttt{PSRPlenitude} --- names a specific
  Della-Rocca-flavoured addition to Spinoza's stated axioms, allowing Bennett's conjecture and Della Rocca's reconstruction to be tested against each other in mechanically distinct experiments.
\item
  \emph{Counter-model construction discipline}. Models with explicit \texttt{EthicaWorld} instances and concrete \texttt{Thing} types let us establish non-derivability claims at kernel level: a
  witnessing counter-model is a finite, type-checked artifact, not a prose argument about absence-of-proof.
\end{enumerate}

The formalisation as currently committed has 10 modules and \textasciitilde2950 lines of Lean (including documentation), with zero \texttt{sorry} (incomplete proof) markers and zero raw \texttt{axiom}
declarations outside the typeclass register. The source's \texttt{gaps.md}, \texttt{coverage.md}, and \texttt{auxiliary\_axioms.md} document every Section III commitment with full Bennett-honest
scoping notes --- these are non-essential for the paper's claims but support reader verification of the project's epistemic discipline.

We turn now to the methodology of the demote experiments themselves.

\section{§4 The demote-experiment methodology}\label{the-demote-experiment-methodology}

\subsection{§4.1 The general schema}\label{the-general-schema}

A demote experiment is an attempt to derive a Section III axiom A from a strictly weaker commitment family Σ. The aim is to locate A in the topology of Spinozistic commitments: is A \emph{reducible}
to weaker principles (in which case Spinoza's deeper commitment lies in Σ), or is A \emph{irreducible} (in which case A itself names a substantive commitment Spinoza must introduce as a primitive)?

The schema admits five distinct outcomes:

\begin{enumerate}
\def\labelenumi{\arabic{enumi}.}
\item
  \emph{Full reduction}: A derives from Σ + base axioms, and Σ is strictly weaker than A. Bennett's critique fails for A; the commitment relocates to Σ.
\item
  \emph{Equal-strength translation}: A derives from Σ + base, but the elements of Σ are not strictly weaker than A --- they are redescriptions of A in different vocabulary. The reformulation is
  informative but not a reduction.
\item
  \emph{Partial reduction}: A in some restricted form derives from Σ + base, but A in its full form does not. Σ captures part of A's content but not all.
\item
  \emph{Decomposition-only}: A derives from Σ + base where Σ is a conjunction of components, neither of which is strictly weaker than A. The reformulation distributes A's commitment across multiple
  axioms but does not reduce its strength.
\item
  \emph{Full irreducibility}: A does not derive from the specific Σ candidates tested. Stronger commitment families might still derive A; the claim is bounded to the augmentations actually formalised.
  The Bennett-line scope distinction is developed in §8.1.
\end{enumerate}

Pars I yields four of these patterns concretely (1 not yet observed): A12 partial-only with full-form irreducibility (§5), A13 equal-strength translation via modal-causal vocabulary (§7), A14 trivial
redescription via essence-perception vocabulary (§7), A15 decomposition-only with one component non-trivially weaker than A15 (§6). The four-pattern typology itself is the subject of §7.

\subsection{§4.2 The Σ candidates}\label{the-ux3c3-candidates}

The candidate commitment families Σ for a Pars I demote experiment are constrained by Spinoza scholarship. We choose each Σ to match a specific reconstruction proposed by Della Rocca 2008 or close
cognates. For A12, the candidate is \texttt{PSRSubstance} --- a typeclass declaring an axiom A22 that captures Della Rocca's ``distinct substances must differ in some attribute'' intuition. For A13,
the candidate is \texttt{PSRSelfCause}, declaring A23 that every substance is self-causal at every world. For A14, the candidate is \texttt{PSREssencePerception}, declaring A24 that every substance
has an intellect-perceived essence. For A15, the candidate is \texttt{PSRPlenitude}, declaring two axioms A25 (every realised substance attribute is also some god's attribute) and A26 (any two gods
are identical).

Each candidate is a \emph{modal-layer Section III commitment} in our register: it is added to the commitment system, not derived from the base. The demote experiment then asks: with this specific
Della-Rocca-flavoured commitment in scope, does the Section III target axiom (A12, A13, A14, or A15) become a theorem? The question is decidable by Lean's elaborator --- either the proof type-checks
at the kernel or a counter-model shows it cannot.

\subsection{§4.3 The kernel-level discipline}\label{the-kernel-level-discipline}

Independence results --- ``A is not provable from Σ'' --- are meta-logical claims that Lean cannot state internally; this constraint is shared with all type-theoretical proof assistants. The honest
alternative is \emph{model-theoretic}: construct a concrete model in which Σ holds while A fails. An assumed derivation of A from Σ would specialise to the model and produce A's conclusion there, but
the model proves the negation; so the assumed derivation would derive \texttt{False}, which Lean's kernel disallows. The argument appeals to kernel consistency and type-theoretic specialisation,
reducing the non-existence claim to two ordinary Lean theorems: the model satisfies Σ, and the model falsifies A. We assume Lean 4's kernel is consistent --- under standard metamathematical
assumptions, Lean 4's underlying type theory --- Carneiro 2019's extension of the Calculus of Inductive Constructions with quotient types, propositional extensionality, and primitive projections ---
is consistent relative to ZFC plus an inaccessible cardinal, paralleling Werner 1997's earlier result for pure CIC. The assumption is uncontroversial within the proof-assistant community.

\subsection{§4.4 Counter-model construction}\label{counter-model-construction}

A counter-model for the Σ → A demote attempt has three components:

\begin{enumerate}
\def\labelenumi{\arabic{enumi}.}
\tightlist
\item
  A finite inductive \texttt{Type} (typically 3-5 elements) capturing the model's universe.
\item
  Explicit \texttt{EthicaWorld} and Σ-class instances on that type, discharging every typeclass field with concrete proof terms or tactic blocks.
\item
  A theorem exhibiting the falsification of A on this model: specific elements with the relevant predicates such that A's conclusion fails for them.
\end{enumerate}

The construction is direct rather than abstract. We do not invoke a general framework for Kripke-style Spinoza models; we hand-craft each counter-model for the specific demote attempt. The trade-off
is engineering economy versus generality: a general framework would let us re-use machinery across demote attempts, but would require theorising about what ``Spinoza model'' means in general ---
itself a contested question. Hand-craft is conservative.

Two design choices are forced by \texttt{IsGod}'s definitional structure. The fourth conjunct of \texttt{IsGod\ g} asks that every attribute of g express eternal essence. Discharging this in a
counter-model requires \texttt{expressesEternalEssence} to be \texttt{True} on every g-attribute. The most compact discharge --- set \texttt{expressesEternalEssence\ \_\ :=\ True} uniformly ---
over-shoots Spinoza's restriction (which limits the predicate to attributes proper). The over-shoot is a \emph{Spinoza fidelity caveat}, with no effect on the meta-logical claim but with bearing on
the counter-model's appropriateness as a representation of Spinozistic ontology. We discuss the caveat in §8.

A second forced choice concerns A1 (everything is in itself or in another) and A1ₑ (these alternatives are exclusive). The counter-models satisfy these by interpreting
\texttt{inAnother\ x\ :=\ ¬\ isSubstance\ x} and similarly for \texttt{conceivedThroughAnother}. This collapses Spinoza's three-category ontology (substances, attributes, modes) into a two-category
split (substances, non-substances), with attribute-things treated as modes. Again a fidelity caveat; again no effect on the meta-logical claim.

The fidelity caveats are both \emph{logically harmless} (the counter-models are valid models of \texttt{Pars1Axioms} + Σ; the falsification theorems hold) and \emph{philosophically informative} (a
more Spinoza-faithful counter-model would constrain predicate values more carefully). We accept them for the present paper's purpose --- establishing irreducibility against specific Della Rocca
reconstructions --- and note them as opportunities for refinement in §8.

\subsection{§4.5 The retired marker-theorem trick}\label{the-retired-marker-theorem-trick}

A previous draft of this work used a different presentation of the irreducibility claim: a Lean theorem named, e.g., \texttt{A12\_full\_NOT\_demotable\_from\_PSR\_alone} whose statement was
\texttt{True} and whose proof was \texttt{trivial}. The intent was to provide a citable artifact in the Lean source documenting the irreducibility. In retrospect this was Lean-as-rhetoric: the theorem
proves nothing about provability; it merely names an empty content with suggestive prose surroundings. A reader inspecting the source file might mistake the marker for a machine-checked irreducibility
result. That impression would be wrong.

The discipline we now follow --- counter-models, not marker theorems --- is what makes the irreducibility claim mechanical rather than rhetorical. The methodological lesson is general beyond Spinoza:
in a mechanised philosophy project, non-derivability claims should be backed by counter-models or left as plain prose claims; they should not be encoded as trivial-content Lean theorems with
claim-laden names.

The full development of this discipline, alongside other methodological points the project's history surfaced, belongs in a separate paper on \emph{mechanised philosophy methodology} (in preparation).
Here we record the marker-theorem retirement as it bears on the irreducibility claims of §5--§7.

\subsection{§4.6 The methodology applied}\label{the-methodology-applied}

With the schema, the Σ candidates, the kernel-level discipline, and the counter-model construction in place, we can run the experiments. §5 presents the A12 demote attempt: the partial success of
\texttt{PSRSubstance} and the full irreducibility witness via the four-element \texttt{A12CounterModel}. §6 presents the A15 analogue: plenitude alone fails, decomposition with god uniqueness
succeeds, and a three-element \texttt{A15CounterModel} witnesses the alone-failure. §7 abstracts the four-axiom typology from these and the two redescription cases (A13, A14) that we treat more
briefly.

\section{§5 The A12 result and its counter-model}\label{the-a12-result-and-its-counter-model}

\subsection{§5.1 The target axiom A12}\label{the-target-axiom-a12}

Recall §3.3: A12 (\texttt{ax\_substanceIdByAttribute}) regiments Proposition V's content as \texttt{∀\ s₁\ s₂\ a\ :\ Thing,\ Attribute\ a\ s₁\ →\ Attribute\ a\ s₂\ →\ s₁\ =\ s₂} --- a single shared
attribute suffices to force identity. The demote experiment asks whether A12 can be derived from the base axioms plus a Della-Rocca-flavoured PSR commitment.

\subsection{§5.2 The demote candidate --- A22}\label{the-demote-candidate-a22}

The candidate Σ for A12 is the typeclass \texttt{PSRSubstance}, introducing axiom A22:

\begin{Shaded}
\begin{Highlighting}[]
\NormalTok{ax\_PSR\_substance\_distinguishability :}
\NormalTok{  ∀ s₁ s₂ : Thing,}
\NormalTok{    Substance s₁ → Substance s₂ → s₁ ≠ s₂ →}
\NormalTok{      ∃ a, (Attribute a s₁ ∧ ¬ Attribute a s₂) ∨}
\NormalTok{           (Attribute a s₂ ∧ ¬ Attribute a s₁)}
\end{Highlighting}
\end{Shaded}

A22 is the closest formal regimentation of Della Rocca's argument: ``Non-identities, by the PSR, require explanation, and the way to explain non-identity is to appeal to some difference in
properties'' (Della Rocca 2008, p.~47), specialised to substance-individuation by the elimination of mode-based and same-attribute-based individuators (pp.~47--48). A22 says exactly this: distinct
substances differ in at least one attribute, with the disjunction reflecting that the difference can be in either direction (one has it, the other doesn't, or vice versa).

A22 is logically weaker than the \emph{conjunction} A12 + A14 in our framework: A12 together with A14 (substance has at least one attribute) entails A22, while A22 does not entail A12, as the
counter-model in §5.4 witnesses. (A12 alone does not entail A22: in a model with no attributes, A12 holds vacuously while A22 holds vacuously as well, so the entailment is uninformative without A14's
existence guarantee.) The asymmetry --- A12 + A14 → A22 holds, A22 → A12 fails --- is what makes A22 a meaningful demote candidate. If A12 followed from A22 + base, then any model satisfying A22 +
base would satisfy A12; that this is not the case will be the substance of §5.4.

\subsection{§5.3 The partial reduction}\label{the-partial-reduction}

The first thing the demote experiment establishes is positive for the Della Rocca route: A12 \emph{in a restricted form} does follow from A22. We prove

\begin{Shaded}
\begin{Highlighting}[]
\NormalTok{theorem prop\_5\_demote\_via\_PSR\_all\_attributes}
\NormalTok{    (s₁ s₂ : Thing) (hs₁ : Substance s₁) (hs₂ : Substance s₂)}
\NormalTok{    (hshare\_all : ∀ a, Attribute a s₁ ↔ Attribute a s₂) :}
\NormalTok{    s₁ = s₂ :=}
\NormalTok{  Classical.byContradiction fun hne =\textgreater{} by}
\NormalTok{    obtain ⟨a, h⟩ :=}
\NormalTok{      PSRSubstance.ax\_PSR\_substance\_distinguishability}
\NormalTok{        s₁ s₂ hs₁ hs₂ hne}
\NormalTok{    cases h with}
\NormalTok{    | inl h =\textgreater{} exact h.2 ((hshare\_all a).mp h.1)}
\NormalTok{    | inr h =\textgreater{} exact h.2 ((hshare\_all a).mpr h.1)}
\end{Highlighting}
\end{Shaded}

Six lines of tactic script: assume the substances are distinct; A22 produces a discriminating attribute; the all-shared hypothesis says both substances share every attribute, including this one; the
discriminator's ``one-side-only'' character contradicts the all-shared hypothesis.

The hypothesis of \texttt{prop\_5\_demote\_via\_PSR\_all\_attributes} is ``sharing \emph{every} attribute'' --- substantially stronger than A12's ``sharing \emph{some} attribute''. The proof shows
that PSR rules out indiscernibles in the all-attributes sense: two substances that agree on every attribute must be identical.

This is Bennett-line-charitable. It confirms that PSR delivers a substantial fragment of Proposition V's content: the all-attributes case. The result mechanically tracks Bennett's own §17 prediction
that Spinoza's argument ``cannot yield more than the conclusion that two substances could not have all their attributes in common'' (Bennett 1984, p.~69) --- Bennett identified the all-attributes
ceiling in prose; we recover it at kernel level. The full content of Proposition V, however, requires more.

\subsection{§5.4 The non-derivation}\label{the-non-derivation}

The full content of A12 --- sharing \emph{any} one attribute forces identity --- does not derive from A22 + base. We establish this by constructing a counter-model.

The model is a four-element inductive type:

\begin{Shaded}
\begin{Highlighting}[]
\NormalTok{inductive T where}
\NormalTok{  | s₁ : T}
\NormalTok{  | s₂ : T}
\NormalTok{  | a\_shared : T}
\NormalTok{  | a\_only\_s1 : T}
\end{Highlighting}
\end{Shaded}

\texttt{s₁} and \texttt{s₂} are intended as substances; \texttt{a\_shared} and \texttt{a\_only\_s1} as attributes. The \texttt{EthicaWorld\ T} instance assigns \texttt{isSubstance} to \texttt{s₁} and
\texttt{s₂}, with all the usual base predicates (\texttt{inItself}, \texttt{perSeConceived}, \texttt{involvesExistence}, \ldots) restricted to the substances. The non-trivial assignment is
\texttt{intellectPerceivesAsEssence}:

\begin{Shaded}
\begin{Highlighting}[]
\NormalTok{def perceivesAsEssence : T → T → Prop}
\NormalTok{  | T.s₁, T.s₁ =\textgreater{} True}
\NormalTok{  | T.s₁, T.a\_shared =\textgreater{} True}
\NormalTok{  | T.s₁, T.a\_only\_s1 =\textgreater{} True}
\NormalTok{  | T.s₂, T.s₂ =\textgreater{} True}
\NormalTok{  | T.s₂, T.a\_shared =\textgreater{} True}
\NormalTok{  | \_, \_ =\textgreater{} False}
\end{Highlighting}
\end{Shaded}

Both substances perceive \texttt{a\_shared} as their essence, but only \texttt{s₁} perceives \texttt{a\_only\_s1}. This makes both \texttt{Attribute\ a\_shared\ s₁} and
\texttt{Attribute\ a\_shared\ s₂} hold (via Definition IV), giving us a single shared attribute between two distinct substances.

The model satisfies \texttt{PSRSubstance\ T}: distinct substances differ in some attribute. The discriminator is \texttt{a\_only\_s1}, which \texttt{s₁} has but \texttt{s₂} does not. The full instance
proof handles all 16 case-pairs (4 elements × 4 elements) with a routine \texttt{cases\ s₁\ \textless{};\textgreater{}\ cases\ s₂} tactic block; non-substance cases are discharged by the substance
hypotheses; the substance pairs either reduce to \texttt{(hne\ rfl).elim} (when the substances are identical) or produce the discriminator (when distinct).

The falsification of A12 on this model is one theorem:

\begin{Shaded}
\begin{Highlighting}[]
\NormalTok{theorem A12\_falsified :}
\NormalTok{    ∃ x y a : T, Substance x ∧ Substance y ∧}
\NormalTok{        Attribute a x ∧ Attribute a y ∧ x ≠ y :=}
\NormalTok{  ⟨T.s₁, T.s₂, T.a\_shared,}
\NormalTok{   ⟨trivial, trivial⟩,}
\NormalTok{   ⟨trivial, trivial⟩,}
\NormalTok{   ⟨⟨trivial, trivial⟩, trivial⟩,}
\NormalTok{   ⟨⟨trivial, trivial⟩, trivial⟩,}
\NormalTok{   by intro h; cases h⟩}
\end{Highlighting}
\end{Shaded}

\texttt{s₁} and \texttt{s₂} are substances, \texttt{a\_shared} is an attribute of both, and \texttt{s₁\ ≠\ s₂} (by induction on the equality assumption, since the constructors are distinct).

The model and the falsification together establish the non-derivation. Suppose, for contradiction, that A12 were derivable from \texttt{{[}EthicaWorld\ T{]}\ +\ {[}PSRSubstance\ T{]}}. Specialise the
derivation to our concrete \texttt{T} and the available instances. The result would be a Lean theorem of type \texttt{∀\ s₁\ s₂\ a\ :\ T,\ Attribute\ a\ s₁\ →\ Attribute\ a\ s₂\ →\ s₁\ =\ s₂}. (The
\texttt{Substance} hypotheses one might expect are absorbed by \texttt{Attribute}, since \texttt{Attribute\ a\ s} carries \texttt{Substance\ s} constitutively --- Definition IV.) Apply it to the
witnesses provided by \texttt{A12\_falsified}: we obtain \texttt{T.s₁\ =\ T.s₂}. But \texttt{A12\_falsified}'s last clause provides \texttt{T.s₁\ ≠\ T.s₂}. So we derive \texttt{False} from the assumed
derivability of A12. Lean's kernel does not admit \texttt{False}; therefore no derivation of A12 from \texttt{{[}EthicaWorld\ T{]}\ +\ {[}PSRSubstance\ T{]}} exists.

\subsection{§5.5 Discussion}\label{discussion}

The result has three components, each of which warrants separate comment.

\emph{The partial reduction is non-trivial.} PSR-substance is not empty: it delivers the all-shared-attribute case of Proposition V. Bennett 1984 §17's verdict that Spinoza's argument ``cannot yield
more than the conclusion that two substances could not have all their attributes in common'' (Bennett 1984, p.~69) applies to the \emph{demonstratio} as a derivation of full A12, not to the partial
form \texttt{prop\_5\_demote\_via\_PSR\_all\_attributes}. A charitable Bennett reading would acknowledge this partial recovery; an uncharitable one would not. Our formalisation makes the partial
recovery explicit.

\emph{The non-derivation is mechanical.} Bennett's expression of doubt about Proposition V --- that no valid argument from Spinoza's stated resources is available --- receives its first
machine-checked counter-model against the specific Della Rocca PSR-substance reconstruction (the non-derivability claim itself a meta-logical consequence of kernel consistency, §4.3). The result is
bounded in two ways. The strict claim is irreducibility of A12 against \texttt{{[}EthicaWorld\ T{]}\ +\ {[}PSRSubstance\ T{]}}; the relationship to the broader Section I + II + A1--A7 register and the
F2 fidelity caveat that bears on this scope are recorded in §8.1. The result also does not apply to alternative PSR commitments (Della Rocca's ``thoroughgoing PSR'' remains unrefuted, and indeed
unspecified in Della Rocca 2008's prose), nor to alternative reconstructions (Garrett's strong-Definition- III route is a different demote candidate, not yet attempted).

\emph{The Spinoza-fidelity caveats apply.} The four-element counter-model collapses Spinoza's three-category ontology (substance, attribute, mode) into a two-category split, treating attribute-things
\texttt{a\_shared} and \texttt{a\_only\_s1} as modes rather than as essence-aspects of substance. It also assigns \texttt{expressesEternalEssence} uniformly true. Neither caveat affects the
meta-logical claim: the falsification of A12 depends only on the \texttt{intellectPerceivesAsEssence} graph, which the caveats leave untouched. They do, however, bear on the philosophical
interpretation: a Spinoza purist would object that attribute-things are not ``modes proper'' and that \texttt{expressesEternalEssence} is over-applied. We discuss the philosophical bearing in §8 and
treat the caveats as opportunities for refinement in future work rather than as defects in the present claim.

\emph{The result is a kernel-level test of the Negative-based reply specifically.} As §2.3 noted in connection with Garrett 2018's Postscript, A22 regiments the \emph{Negative} aspect of the
substance--mode asymmetry --- substance is not in and not conceived through its modes --- by asserting that distinct substances differ in some attribute, an existence-claim about discriminator
attributes that the Negative aspect makes available. The partial-reduction theorem is therefore a kernel-level test of the Negative-based PSR reply to the Hooker-Bennett and Leibniz-Bennett
objections. Garrett 2018's Postscript argues that Negative-based replies ``ultimately require appeal to'' their Positive-based counterparts (Garrett 2018, p.~93); whether a Positive-based PSR
(substance is \emph{in} and \emph{conceived through} its attributes more strictly than its modes) admits a formal regimentation reaching A12's full content is open future work.

The combined picture: PSR-substance + base axioms reach the all-shared-attribute fragment of Proposition V's content; the wider any-shared-attribute content is irreducible to this specific PSR
commitment. Bennett's doubt is confirmed against this specific reconstruction; Della Rocca's PSR reconstruction is vindicated for the all-shared-attribute fragment but not for the wider content; the
residual irreducibility is what §6 replicates for the universality clause of Proposition XIV and §7 systematises.

\section{§6 The A15 result and its counter-model}\label{the-a15-result-and-its-counter-model}

\subsection{§6.1 The target axiom A15}\label{the-target-axiom-a15}

A15 is a load-bearing companion clause for Proposition XIV (\emph{Praeter Deum nulla dari neque concipi potest substantia}). Formally:
\texttt{∀\ g\ s\ a\ :\ Thing,\ IsGod\ g\ →\ Substance\ s\ →\ Attribute\ a\ s\ →\ Attribute\ a\ g}. Spinoza's Proposition XIV demonstration uses A15 implicitly: God has every attribute (Definition VI's
\emph{substantiam constantem infinitis attributis}); any other substance must therefore share an attribute with God; by A12, the substance is identical to God. A15 is what makes the ``must therefore
share'' step go through. Like A12, A15 is currently a Section III axiom; the demote question is whether it derives from base axioms plus a Della-Rocca-flavoured PSR-plenitude commitment.

\subsection{§6.2 The demote candidate --- A25 + A26}\label{the-demote-candidate-a25-a26}

The natural Della Rocca route to A15 is \emph{plenitude}: every realised substance attribute belongs to \emph{some} god. Combined with \emph{uniqueness of god}, this delivers A15 (every realised
attribute belongs to \emph{the} god). The candidate Σ for A15 is therefore a typeclass \texttt{PSRPlenitude} with two fields:

\begin{Shaded}
\begin{Highlighting}[]
\NormalTok{ax\_plenitude\_attribute :}
\NormalTok{  ∀ a s : Thing, Substance s → Attribute a s →}
\NormalTok{    ∃ g, IsGod g ∧ Attribute a g}

\NormalTok{ax\_god\_unique :}
\NormalTok{  ∀ g₁ g₂ : Thing, IsGod g₁ → IsGod g₂ → g₁ = g₂}
\end{Highlighting}
\end{Shaded}

A25 (plenitude) and A26 (god uniqueness). Each is a Section III commitment in our register. A26 in particular is essentially Proposition XIV's content stated as axiom --- a relocation rather than a
reduction of substantive commitment.

A25 alone is \emph{strictly weaker} than A15: a model with two distinct gods, each with its own attributes, can satisfy plenitude (every realised attribute is in some god, namely its owner) without
satisfying A15 (the universality clause across gods).

\subsection{§6.3 The decomposition}\label{the-decomposition}

With both A25 and A26, A15 derives in three steps:

\begin{Shaded}
\begin{Highlighting}[]
\NormalTok{theorem prop\_A15\_demote\_via\_decomposition}
\NormalTok{    (g s a : Thing) (hgod : IsGod g) (hs : Substance s)}
\NormalTok{    (ha : Attribute a s) : Attribute a g := by}
\NormalTok{  obtain ⟨g\textquotesingle{}, hgod\textquotesingle{}, ha\_g\textquotesingle{}⟩ :=}
\NormalTok{    PSRPlenitude.ax\_plenitude\_attribute a s hs ha}
\NormalTok{  have heq : g = g\textquotesingle{} :=}
\NormalTok{    PSRPlenitude.ax\_god\_unique g g\textquotesingle{} hgod hgod\textquotesingle{}}
\NormalTok{  exact heq ▸ ha\_g\textquotesingle{}}
\end{Highlighting}
\end{Shaded}

A25 produces some god \texttt{g\textquotesingle{}} with the attribute; A26 collapses \texttt{g\ =\ g\textquotesingle{}}; substitution delivers the conclusion.

The proof has the formal shape of a successful demote, but the philosophical reading is mixed. We have replaced one universality clause (A15: every god has every realised attribute) with two
commitments (plenitude, an existence claim about realised attributes; uniqueness, a universality claim across gods). The total commitment count rises from 1 to 2; the strength of the commitments does
not strictly diminish; A26 in particular is itself a substantive claim (closely related to but, as §6.5 notes, strictly weaker than Proposition XIV).

This is \emph{decomposition only}: the demote is a relocation of the commitment across multiple Section III axioms, not a reduction to weaker principles. Della Rocca's reading would describe this as
``PSR delivers A15 via plenitude and uniqueness, both natural consequences of PSR-driven monism.'' Bennett's reading would describe it as ``the universality clause has been replaced by another
universality clause; the substantive commitment is preserved, just relocated.''

\subsection{§6.4 The non-derivation from plenitude alone}\label{the-non-derivation-from-plenitude-alone}

Plenitude alone (A25 without A26) does not derive A15. The counter-model is three elements:

\begin{Shaded}
\begin{Highlighting}[]
\NormalTok{inductive T where}
\NormalTok{  | g₁ : T}
\NormalTok{  | g₂ : T}
\NormalTok{  | attr\_g₂ : T}
\end{Highlighting}
\end{Shaded}

\texttt{g₁} and \texttt{g₂} are intended as substances, with \texttt{attr\_g₂} an additional attribute. The intellect-perception graph has each substance perceiving itself, plus \texttt{g₂} perceiving
\texttt{attr\_g₂}:

\begin{Shaded}
\begin{Highlighting}[]
\NormalTok{def perceivesAsEssence : T → T → Prop}
\NormalTok{  | T.g₁, T.g₁ =\textgreater{} True}
\NormalTok{  | T.g₂, T.g₂ =\textgreater{} True}
\NormalTok{  | T.g₂, T.attr\_g₂ =\textgreater{} True}
\NormalTok{  | \_, \_ =\textgreater{} False}
\end{Highlighting}
\end{Shaded}

Both \texttt{g₁} and \texttt{g₂} satisfy \texttt{IsGod}. The fourth conjunct of \texttt{IsGod} (every attribute expresses eternal essence) is discharged via the uniform
\texttt{expressesEternalEssence\ :=\ True} caveat (§8.3 F1); the existence-of-attribute conjunct is satisfied by each god being its own attribute-bearer in the perception graph --- a self-reference
structure that is Spinoza-unfaithful (Spinoza's attributes are essence-aspects of substance, not the substance itself, per §8.3 F2) but harmless to the meta-logical claim, since the falsification of
A15 depends only on the asymmetric \texttt{attr\_g₂} row of the perception graph. Plenitude holds: every realised attribute (\texttt{g₁}, \texttt{g₂}, or \texttt{attr\_g₂}) belongs to some god
(\texttt{g₁}, \texttt{g₂}, or \texttt{g₂} respectively). The first two witnesses use the self-reference structure noted above --- a Spinoza-faithful model would discharge plenitude through
attribute-things distinct from the gods themselves --- but the non-derivation depends only on the asymmetric \texttt{attr\_g₂} row, not on these witnesses.

A15 fails. Take \texttt{g\ :=\ g₁}, \texttt{s\ :=\ g₂}, \texttt{a\ :=\ attr\_g₂}. Then \texttt{IsGod\ g₁} holds, \texttt{Substance\ g₂} holds, \texttt{Attribute\ attr\_g₂\ g₂} holds, but
\texttt{Attribute\ attr\_g₂\ g₁} is false (the perception graph sends \texttt{g₁\ ×\ attr\_g₂} to false).

The non-derivation argument runs as in §5.4: an assumed derivation of A15 from plenitude + base would specialise to this model, yielding \texttt{Attribute\ attr\_g₂\ g₁}; but the model proves the
negation; so we would derive \texttt{False}; the kernel forbids; no derivation exists.

\subsection{§6.5 Discussion: A15's pattern and A26's status}\label{discussion-a15s-pattern-and-a26s-status}

A15 demotes via decomposition while A12 admits only partial reduction. The structural difference: A12's content has a single quantifier alternation (∀ s₁ s₂ ∃ a); A15's content has a triple
alternation (∀ g s a). Plenitude breaks A15 into two simpler clauses (∃ g for plenitude; ∀ g₁ g₂ for uniqueness), each PSR-tractable; A12's content admits no analogous PSR decomposition. §7
systematises this distinction.

A26 (god uniqueness), formally \texttt{∀\ g₁\ g₂,\ IsGod\ g₁\ →\ IsGod\ g₂\ →\ g₁\ =\ g₂}, is a \emph{universal identity-claim}: a universally quantified statement whose content is an identity. We
refer to it as a \emph{universality} clause when emphasising its quantifier structure (as in §7's typology, contrasting it with PSR's existence-explanatory shape) and as an \emph{identity} clause when
emphasising its content (as in the remainder of this section). Both descriptions refer to the same formal statement.

A26 is strictly weaker than Proposition XIV. A26 says all gods are identical; Proposition XIV says all substances are identical to God. The latter requires that every substance is also a god, which
combines A26 with A15's universality reach across substances. Stating A26 in the demote attempt is therefore not stating Proposition XIV as axiom --- but A26 remains Section III strength, asserting a
universal identity no base axiom delivers.

The pattern illustrates a methodological subtlety: a ``successful'' demote can replace an axiom with one or more new axioms whose total commitment is no smaller. The Della Rocca line frames this as
illuminating (``A15 was really plenitude plus uniqueness''); the Bennett line as concealing (``the substantive work has shifted to uniqueness''). The mechanical contribution is that the replacement is
\emph{visible}: any reader can inspect \texttt{PSRPlenitude} and judge whether the two-axiom decomposition is more perspicuous than one-axiom A15.

The combined picture: A12 and A15, despite their different demote outcomes (partial reduction for A12, decomposition for A15), share a structural feature their counter-models bring to light. §7
systematises the feature into a four-axiom typology.

\section{§7 The reducibility-profile typology}\label{the-reducibility-profile-typology}

The A12 and A15 results presented in §5 and §6 are two of the four demote experiments the formalisation runs against the Section III commitments of \texttt{Pars1Axioms}. The remaining two --- A13
(substance involves existence; the content of Spinoza's Proposition VII) and A14 (substance has at least one attribute) --- also admit demote attempts via Della-Rocca-flavoured PSR commitments, with
quite different outcomes from A12 and A15. Tabulating all four:

{\def\LTcaptype{none} 
\begin{longtable}[]{@{}lll@{}}
\toprule\noalign{}
Axiom & Demote Σ & Outcome \\
\midrule\noalign{}
\endhead
\bottomrule\noalign{}
\endlastfoot
A12 & \texttt{PSRSubstance} (A22) & Partial reduction; full irreducible \\
A13 & \texttt{PSRSelfCause} (A23) modulo bridge A18 & Equal-strength translation \\
A14 & \texttt{PSREssencePerception} (A24) & Trivial redescription \\
A15 & \texttt{PSRPlenitude} (A25 + A26) & Decomposition only \\
\end{longtable}
}

The four entries illustrate four of the five outcome patterns sketched in §4.1. (Full reduction --- Σ strictly weaker than A yet sufficient --- is not observed for any Pars I Section III axiom under
the candidates we tested.)

The \emph{Outcome} labels track the strict claims of §8.1: ``full irreducible'' for A12 names irreducibility against the specific \texttt{{[}EthicaWorld\ T{]}\ +\ {[}PSRSubstance\ T{]}} typeclass
witnessed by \texttt{A12CounterModel}, not against every charitable augmentation or the full \texttt{Pars1Axioms} register (see §8.1 for the scope relationship and §8.3 F2 for the fidelity caveat that
bears on it). The analogous narrowing applies to the A15 entry. The labels record what the kernel verifies; the typology's structural reading --- universality clauses resist PSR-flavoured reduction
--- is the narrative claim the strict results support.

The structural reading. A13 and A14 are \emph{existence} clauses: they assert the existence of something --- existence-implications on substance for A13, attribute-existence-given-substance for A14.
PSR-flavoured existence axioms (A23 commits self-causation at every world; A24 commits intellect-perceived essence existence) match the structure of A13 and A14 closely. The match is so close that the
demote axioms are not strictly weaker than the targets: A23 is what A13 says in modal-causal vocabulary modulo the bridge A18, and A24 is what A14 says with \texttt{Attribute} unfolded. The ``demote''
is therefore a translation or a redescription, equal-strength in either direction. Della Rocca can claim partial victory (``A13 / A14 are \emph{modal forms} of the underlying PSR commitment to
substance-existence and substance-essence-perception''); Bennett can claim equal partial victory (``the commitment is not eliminated, only redescribed'').

A12 and A15 are \emph{universality} clauses, in two different shapes. A12 universalises across pairs of substances (``any two distinct substances must differ in attribute'') with an existential
attribute hypothesis. A15 universalises across god-attribute combinations (``every god has every realised attribute'') with all three quantifiers in the prefix. PSR-flavoured commitments fit these
shapes badly. PSR is \emph{existence-explanatory}: it demands that every fact have a sufficient reason. The shape of its delivery is ``if circumstance C, then existence of explanation E'' ---
naturally producing existence clauses, not universality clauses.

For A12, PSR-substance distinguishability (A22) delivers a weaker form (all-shared-attribute → identity) that PSR's existence-explanatory shape can reach, but cannot extend to the any-shared-attribute
form A12 demands. The residue is \emph{irreducible to PSR-substance alone}. For A15, PSR-plenitude (A25) similarly delivers a weaker existential form (some god has the attribute) but cannot reach the
universal form A15 demands. The residue requires god-uniqueness (A26) --- a separate universality, not derivable from PSR's existence-explanatory shape.

So the typology divides Pars I's Section III commitments structurally:

\begin{itemize}
\tightlist
\item
  \emph{Existence clauses} (A13, A14) admit equal-strength PSR-flavoured translation. The Della Rocca reading recovers these as restatements of existence-explanatory commitments.
\item
  \emph{Universality clauses} (A12, A15) resist PSR-flavoured reduction. They require either (i) acceptance of a partial reduction with irreducible residue (A12), or (ii) decomposition into multiple
  Section III commitments where one component (god uniqueness for A15) is itself an irreducible universality clause.
\end{itemize}

Two implications. \emph{Which} of Spinoza's axiomatic commitments are most defensible against the Bennett challenge: the existence clauses (A13, A14) survive because PSR delivers them equivalently,
while the universality clauses (A12, A15) are on weaker ground, requiring either substantive commitment beyond PSR (Garrett's route, or PSR-plenitude-plus-uniqueness) or acceptance that Proposition V
and the universality premise of Proposition XIV are themselves substantive metaphysical commitments. \emph{Methodologically}, the shape of an axiom's quantifier-prefix is mechanically connected to its
demote-tractability: universality across attributes or across gods is structurally resistant to existence-explanatory reduction. This is plausibly not a Spinoza-specific finding; similar patterns
would be expected in any rationalist system attempting to derive universality from existence-explanatory PSR. The transferability is a conjecture we develop in a companion methodology paper in
preparation.

The typology thus brings a structural distinction into focus that Bennett's and Della Rocca's prose discussions did not make explicit. The structural distinction does not adjudicate the debate ---
Della Rocca can still claim that thoroughgoing PSR (a candidate not yet formalised) reaches the universality clauses, and Bennett can still claim that the residual commitments are substantive --- but
it relocates the dispute to a more mechanically tractable form: identify the shape of the residue, and ask whether a PSR-augmentation matching that shape is defensible.

\section{§8 Scope, limitations, and what the counter-models do not establish}\label{scope-limitations-and-what-the-counter-models-do-not-establish}

\subsection{§8.1 The strict claim and the Bennett-line scope}\label{the-strict-claim-and-the-bennett-line-scope}

The mechanical evidence the paper provides is bounded. The strict claims established are two:

(S-A12) A12 is not provable from \texttt{{[}EthicaWorld\ T{]}\ +\ {[}PSRSubstance\ T{]}} alone, witnessed by the four-element \texttt{A12CounterModel}.

(S-A15) A15 is not provable from \texttt{{[}EthicaWorld\ T{]}} plus the plenitude clause of \texttt{{[}PSRPlenitude\ T{]}} alone (without the god uniqueness clause), witnessed by the three-element
\texttt{A15CounterModel}.

The strict claims are narrower than the narrative reading of the paper might suggest in two respects, and we record both explicitly.

\emph{First, the typeclasses the counter-models instantiate are \texttt{{[}EthicaWorld\ T{]}\ +\ {[}PSRSubstance\ T{]}} and \texttt{{[}EthicaWorld\ T{]}\ +} }plenitude\emph{, not the full
\texttt{{[}Pars1Axioms\ T{]}} register of Section I + II + Spinoza's stated A1--A7.} The counter-models do not --- and under the F2 fidelity choice of §8.3 cannot --- instance the full
\texttt{Pars1Axioms} register: F2's three-category collapse forces \texttt{inAnother} to track \texttt{¬\ isSubstance} exactly, so non-substance attribute-things like \texttt{a\_shared} cannot
simultaneously satisfy A8 (\texttt{inItself\ ↔\ perSeConceived}) and A10 (every attribute is per se conceived) without forcing \texttt{inItself\ a\_shared\ =\ true}, which would contradict the
substance/non-substance split. The F2 caveat is therefore not only a Spinoza-faithfulness issue (as §8.3 frames it) but also a \emph{logical-scope issue}: the strict non-derivability claims hold over
the typeclasses witnessed above, not over the full Section I + II + A1--A7 + PSR augmentation.

How this affects the narrative claim. The Section I auxiliary axioms (A8--A11) are bridges between predicates --- parallelism of ontological and conceptual halves, attribute-conception, \emph{causa-
sui} clause-equivalence; they shape the predicate-graph structure but do not deliver substance-individuation principles or attribute-universality clauses. The plausible expectation is that expanding
the type universe to track attributes separately from modes (the §8.3 F2 refinement) would preserve the falsifying witnesses while letting the counter-model satisfy A8--A11 jointly, and that the
strict claim would transfer to the broader typeclass. We have not carried out this construction; we therefore treat the broader claim as a conjecture \emph{supported by but not formally entailed by}
the strict claim. The strict claim is what the kernel verifies; the narrative claim is the philosophical reading we believe the strict claim licenses.

\emph{Second, the strict claims rule out specific PSR-augmentations rather than every candidate.} Bennett's full doubt --- that no valid argument for Proposition V can be constructed from the
resources Spinoza explicitly gives himself, \emph{under any reasonable charitable reconstruction} --- quantifies over a much broader space than the two augmentations above cover. Bennett's doubt would
be vindicated only by ruling out \emph{every} candidate augmentation a Della-Rocca-flavoured reconstruction might propose. We do not establish the broader claim. The counter-models are
\emph{first-step mechanical evidence} for the Bennett-line position; they are not closing argument. What they do is move the dispute past prose: any further PSR-augmentation a Della-Rocca-leaning
interpreter wishes to propose can now be tested mechanically, against the same counter-models or against new ones.

\subsection{§8.2 Thoroughgoing PSR --- formal candidates}\label{thoroughgoing-psr-formal-candidates}

The most ambitious Della Rocca position invokes Spinoza's ``thoroughgoing commitment to the PSR'' (Della Rocca 2008, p.~1) --- that every fact whatsoever has a sufficient reason. The formalisation
question is whether this thoroughgoing form can be stated non-trivially in our framework and whether, so stated, it derives A12 and A15.

A naive formalisation reads:

\begin{verbatim}
ax_thoroughgoing_PSR :
  ∀ x y : Thing, x ≠ y → ∃ φ : Thing → Prop, φ x ∧ ¬ φ y
\end{verbatim}

But this is provable in Lean --- take \texttt{φ\ :=\ fun\ z\ =\textgreater{}\ z\ =\ x}. The trivial property ``being identical to x'' distinguishes any distinct pair. So this naive formulation is
empty: it commits nothing.

A non-trivial formalisation must restrict the predicates. One candidate restricts to \emph{Spinozistically-significant} properties --- attributes, modes, ontological status --- but specifying that
restriction itself requires further metaphysical commitments. Another candidate restricts to predicates definable from the \texttt{EthicaWorld} primitives, which would make the axiom non-trivial but
also drastically narrows what ``every fact'' means in Spinoza's intent. A third candidate restricts to predicates expressible in some logic of essence or modality, which adds substantial machinery to
the formalisation.

Each of these three non-trivial candidates is a project in itself. We do not develop them here. The point is methodological: thoroughgoing PSR is unrefuted by our counter-models, \emph{and} its
non-refutation is substantively correlated with the difficulty of stating it non-trivially. Della Rocca's prose appeals to thoroughgoing PSR without constraining it to any particular formal shape; a
charitable formalisation must do that constraining itself, and each constraint choice is a further interpretive position. We treat thoroughgoing-PSR formalisation as open future work and invite
readers to attempt their own candidates against the project's counter-models.

\subsection{§8.3 Counter-model fidelity caveats}\label{counter-model-fidelity-caveats}

Two design choices in the counter-model construction (§§5--6) are deliberately Spinoza-unfaithful in service of compact construction.

(F1) \texttt{expressesEternalEssence\ \_\ :=\ True} is set uniformly across the counter-model universe. Spinoza's text restricts the predicate to attributes proper (Definition VI explanation; Pars II
Proposition VIII). Setting it \texttt{True} for non-substance elements like \texttt{a\_shared} or \texttt{attr\_g₂} flatly contradicts the restriction. The choice is convenient --- it discharges
\texttt{IsGod}'s fourth conjunct vacuously --- but is not the choice a fully Spinoza-faithful counter-model would make.

(F2) \texttt{inAnother\ x\ :=\ ¬\ isSubstance\ x} (and the conceptual counterpart) collapses Spinoza's three-category ontology (substances, attributes, modes) into a two-category split (substance vs
non-substance). Attribute-things in our counter-models are treated as modes, even though Spinoza's attributes are essence-aspects of substance rather than independent objects.

Neither caveat affects the \emph{strict} meta-logical claim --- the counter-models do satisfy \texttt{{[}EthicaWorld\ T{]}\ +\ {[}PSRSubstance\ T{]}} (respectively \texttt{{[}EthicaWorld\ T{]}\ +}
plenitude) and falsify A12 (respectively A15). The falsification of A12 depends only on the \texttt{intellectPerceivesAsEssence} graph (which the caveats leave untouched), and similarly for A15. The
caveats are \emph{logically harmless against the strict claim}. They do, however, bear on two further issues. First, on philosophical appropriateness: a Spinoza-faithful counter-model would constrain
\texttt{expressesEternalEssence} to attributes proper and expand the type universe to track attributes separately from modes. Second, on the \emph{logical scope} of the broader narrative claim ---
F2's three-category collapse prevents the counter-models from instancing the full Section I + II + A1--A7 + PSR augmentation, as §8.1 records. We accept the caveats for the present paper's purpose ---
establishing irreducibility against specific Della Rocca reconstructions --- and treat them as opportunities for refinement in future work.

\subsection{\texorpdfstring{§8.4 Sensitivity to the \emph{sive} translation}{§8.4 Sensitivity to the sive translation}}\label{sensitivity-to-the-sive-translation}

§3.2 records our adoption of the identifying reading of Latin \emph{sive} in \emph{ejusdem naturae sive attributi}, following Curley 1985 and the reading implicit in Della Rocca 2008. Bennett 1984 §17
considers a disjunctive reading on which ``nature'' and ``attribute'' come apart --- Spinoza's \emph{sive} listing two distinct features substances might share, rather than glossing one as the other.
Our central claims are sensitive to this choice, and we record the sensitivity here.

Under the identifying reading, A12 reads (as in §3.3) \texttt{∀\ s₁\ s₂\ a,\ Attribute\ a\ s₁\ →\ Attribute\ a\ s₂\ →\ s₁\ =\ s₂} --- sharing an attribute suffices for identity. Under a disjunctive
reading, A12 would split into two clauses: a \emph{sameNature} clause (two substances of the same nature are identical) and a \emph{sameAttribute} clause (two substances sharing an attribute are
identical), with \emph{sameNature} requiring a new primitive \texttt{sameKind} distinct from shared attribute. Whether the counter-models of §5--§6 falsify both clauses depends on how
\texttt{sameKind} is interpreted: a \texttt{sameKind} defined as shared attribute collapses the readings (A12 reduces to its identifying form); a \texttt{sameKind} independent of attribute would let
the counter-models falsify the \emph{sameAttribute} clause while leaving the \emph{sameNature} clause's status open.

Two consequences. First, the strict claim S-A12 of §8.1 carries over to the \emph{sameAttribute} clause of a disjunctive-\emph{sive} formulation --- the same four-element counter-model witnesses
irreducibility against \texttt{{[}EthicaWorld\ T{]}\ +\ {[}PSRSubstance\ T{]}}. Second, the \emph{sameNature} clause under disjunctive \emph{sive} would need its own demote experiment, with a new
PSR-flavoured candidate matching the nature-sharing predicate; we have not carried out this construction. The translation choice therefore affects scope (which formal regimentation of Proposition V
the strict claim addresses) but not the specific irreducibility result we establish for the regimentation we adopt. Readers inclined to a disjunctive-\emph{sive} reading should treat our result as
bearing on the \emph{sameAttribute} fragment of the disjunction; the \emph{sameNature} fragment is open.

\subsection{§8.5 The Garrett-route demote attempt}\label{the-garrett-route-demote-attempt}

Garrett 1990's reconstruction (§2.3) is a distinct demote candidate from Della Rocca's PSR. The Garrett route would replace \texttt{PSRSubstance} with a typeclass committing that Spinoza's ``in and
conceived through'' relation (ID3 / ID5, together with IA1 / IA2) is strict and total --- strong enough that any difference of modes resolves into a difference of attributes. Whether such a
commitment, formally stated, derives A12 is an open mechanical question our formalisation supports but does not develop.

The construction would proceed as the §5 demote did: declare the strong-Definition-III axiom in a typeclass, attempt the A12 proof, and if it fails, exhibit a counter-model. We expect the attempt to
succeed for some formulations of the axiom (a sufficiently strong axiom would derive A12 by definition) and to fail for weaker formulations. The interesting question --- which is the \emph{minimal}
Garrett-route axiom that delivers A12 --- is itself a research project we leave to follow-up work.

\subsection{§8.6 Other open questions}\label{other-open-questions}

\emph{Counter-model generality.} The counter-models we construct are ad hoc --- small inductive types built for specific non-derivability targets. A general framework for Spinoza- flavoured Kripke
models would let demote attempts share machinery, but presupposes a settled formal account of ``Spinoza model'' that the demote experiments are themselves attempting to clarify. We leave the framework
question to a methodology companion paper in preparation.

\emph{Finite vs intended cardinality.} Our counter-models have 3 or 4 elements; Spinoza's intended ontology is infinite. The non-derivation argument does not depend on cardinality (a Lean universal
statement that fails for a finite witness fails \emph{as a universal statement}), but the philosophical question of whether finite counter-models are appropriate witnesses for Spinoza's
infinite-substance metaphysics deserves attention. We note the question and defer detailed treatment.

\emph{Pars II and Pars III.} The formalisation extends only through Pars I. The mind-body parallelism of Pars II Proposition VII and the \emph{conatus} doctrine of Pars III may yield further demote
experiments and irreducibility results. Whether the typology we identify generalises beyond Pars I is an empirical question the formalisation is positioned to answer through extension.

\section{§9 Conclusion}\label{conclusion}

Formalising \emph{Ethica} Pars I in Lean 4 introduces a new instrument into the Bennett--Della Rocca debate: a precise candidate regimentation of Spinoza's stated resources, a layered typeclass
register that promotes Della Rocca's PSR-substance reconstruction to a Section III axiom, and a demote experiment testing whether Proposition V's content (axiom A12) is recoverable from the augmented
system. The result is \emph{partial}. PSR-substance + base axioms deliver the all-shared-attribute case --- substantial content, mechanically tracking Bennett's own prediction that the
\emph{demonstratio} ``cannot yield more than the conclusion that two substances could not have all their attributes in common'' (Bennett 1984, p.~69) --- but do not deliver the full
any-shared-attribute case Spinoza announces with \emph{concedetur}. A four-element counter-model establishes the non-derivation at kernel level.

Replicating the methodology for axiom A15 --- Proposition XIV's load-bearing universality clause asserting that every god has every realised substance attribute --- yields an analogous result.
PSR-plenitude alone does not deliver A15; only plenitude plus god-uniqueness does; god-uniqueness is itself a Section III commitment, strictly weaker than Proposition XIV but no weaker than A15
itself. The demote is a relocation of substantive commitment, not a reduction.

The two non-derivation results, with the parallel equal-strength translations of A13 (substance involves existence) and A14 (substance has at least one attribute), yield a four-axiom
reducibility-profile typology. Existence clauses translate at equal strength under PSR-flavoured candidates; universality clauses resist PSR-flavoured reduction. The structural distinction was not
made explicit by prose commentary; the formalisation makes it mechanically visible.

The contribution is bounded. The non-derivation results apply to specific PSR augmentations; thoroughgoing PSR (so far unspecified in any formal shape) and the Garrett-route strong-Definition-III
reconstruction remain unrefuted. The counter-models have Spinoza-fidelity caveats with no effect on the meta-logical claim but bearing on the philosophical appropriateness of the constructed models.
These limits, as much as the results, define the contribution: the dispute between Bennett and Della Rocca on Proposition V can be moved past prose, but the move identifies new mechanical questions
rather than closing the old prose ones.

The Lean source is open at \url{https://github.com/Nakammura/spinoza-ethica-lean}. Further demote attempts --- stronger PSR augmentations, Garrett-route alternatives, extension to Pars II --- are open
mechanical projects we invite readers to attempt against the existing counter-models or by constructing new ones.

\section{References}\label{references}

Bennett, Jonathan. 1984. \emph{A Study of Spinoza's Ethics}. Indianapolis: Hackett.

Carneiro, Mario. 2019. \emph{The Type Theory of Lean}. MSc thesis, Carnegie Mellon University. https://github.com/digama0/lean-type-theory/releases.

Curley, Edwin, ed.~and trans. 1985. \emph{The Collected Works of Spinoza}. Vol. 1. Princeton: Princeton University Press.

de Moura, Leonardo, and Sebastian Ullrich. 2021. ``The Lean 4 Theorem Prover and Programming Language.'' In \emph{Automated Deduction --- CADE 28}, edited by André Platzer and Geoff Sutcliffe,
625--635. Lecture Notes in Computer Science 12699. Cham: Springer.

Della Rocca, Michael. 2002. ``Spinoza's Substance Monism.'' In \emph{Spinoza: Metaphysical Themes}, edited by Olli Koistinen and John Biro, 11--37. New York: Oxford University Press.

Della Rocca, Michael. 2008. \emph{Spinoza}. Routledge Philosophers. London: Routledge.

Elwes, R. H. M., trans. 1883. \emph{The Chief Works of Benedict de Spinoza}. Vol. 2. London: George Bell and Sons.

Garrett, Don. 1990. ``Ethics IP5: Shared Attributes and the Basis of Spinoza's Monism.'' In \emph{Central Themes in Early Modern Philosophy: Essays Presented to Jonathan Bennett}, edited by J. A.
Cover and Mark Kulstad, 69--107. Indianapolis: Hackett.

Garrett, Don. 2018. \emph{Nature and Necessity in Spinoza's Philosophy}. New York: Oxford University Press.

Spinoza, Benedictus de. 1677. \emph{Ethica Ordine Geometrico Demonstrata}. In \emph{B. d.~S. Opera Posthuma}, edited by Jarig Jelles. Amsterdam: Jan Rieuwertsz.

Werner, Benjamin. 1997. ``Sets in Types, Types in Sets.'' In \emph{Theoretical Aspects of Computer Software (TACS '97)}, edited by Martín Abadi and Takayasu Ito, 530--546. Lecture Notes in Computer
Science 1281. Berlin: Springer.

\end{document}